\documentclass[twocolumn]{aastex62}
\usepackage[utf8]{inputenc}

\shorttitle{An Ensemble of BNNs for Exoplanetary Atmospheric Retrieval}
\shortauthors{2018 NASA FDL Astrobiology Team II}

\usepackage{todonotes}

\graphicspath{ {./images/} }
\usepackage{mathtools}
\usepackage{amsmath}
\usepackage{bbm}
\usepackage{collcell}
\usepackage{hhline}
\usepackage{pgf}
\usepackage{multirow}
\usepackage{courier} 
\usepackage[caption=false]{subfig}
\usepackage{textcomp}

\begin{document}

\title{An Ensemble of Bayesian Neural Networks for Exoplanetary Atmospheric Retrieval}

\correspondingauthor{\newline Adam D. Cobb (machine learning questions) \newline Michael D. Himes (exoplanetary questions)}
\email{acobb@robots.ox.ac.uk, mhimes@knights.ucf.edu}

\author[0000-0003-2868-6983]{Adam D. Cobb}
\thanks{These two authors contributed equally}
\affil{Department of Engineering Science, University of Oxford}

\author[0000-0002-9338-8600]{Michael D. Himes}
\thanks{These two authors contributed equally}
\affil{Planetary Sciences Group, Department of Physics, University of Central Florida}
\collaboration{}
\author[0000-0001-8185-6094]{Frank Soboczenski}
\affiliation{SPHES, King's College London}

\author[0000-0003-0550-3224]{Simone Zorzan}
\affil{ERIN Department, Luxembourg Institute of Science and Technology}

\author[0000-0001-9011-4420]{Molly D. O'Beirne}
\affiliation{Department of Geology and Environmental Science, University of Pittsburgh}

\author[0000-0001-9854-8100]{Atılım Güneş Bayd{\rlap{\.}\i}n}
\affiliation{Department of Engineering Science, University of Oxford}

\author{Yarin Gal}
\affiliation{Department of Computer Science, University of Oxford}

\author[0000-0003-0354-9325]{Shawn D. Domagal-Goldman}
\affiliation{NASA Goddard Space Flight Center, Greenbelt, MD}

\author[0000-0001-6285-267X]{Giada N. Arney}
\affiliation{NASA Goddard Space Flight Center, Greenbelt, MD}

\author[0000-0001-6138-8633]{Daniel Angerhausen}
\affiliation{CSH Fellow, Center for Space and Habitability, University of Bern, Switzerland}
\affiliation{Blue Marble Space Institute of Science, Seattle, United States}
\collaboration{2018 NASA FDL Astrobiology Team II}

\begin{abstract}
\noindent 

Machine learning is now used in many areas of astrophysics, from detecting exoplanets in Kepler transit signals to removing telescope systematics. 
Recent work demonstrated the potential of using machine learning algorithms for atmospheric retrieval by implementing a random forest to perform retrievals in seconds that are consistent with the traditional, computationally-expensive nested-sampling retrieval method.
We expand upon their approach by presenting a new machine learning model, \texttt{plan-net}, based on an ensemble of Bayesian neural networks
that yields more accurate inferences than the random forest for the same data set of synthetic transmission spectra.
We demonstrate that an ensemble provides greater accuracy and more robust uncertainties than a single model.
In addition to being the first to use Bayesian neural networks 
for atmospheric retrieval, we also introduce a new loss function for Bayesian neural networks 
that learns correlations between the model outputs.
Importantly, we show that designing machine learning models to explicitly incorporate domain-specific knowledge both improves performance and provides additional insight by inferring the covariance of the retrieved atmospheric parameters. We apply \texttt{plan-net} to the Hubble Space Telescope Wide Field Camera 3 transmission spectrum for WASP-12b and retrieve an isothermal temperature and water abundance consistent with the literature. We highlight that our method is flexible and can be expanded to higher-resolution spectra and a larger number of atmospheric parameters.

\end{abstract}


\section{INTRODUCTION}\label{sec:intro}

Over a decade ago, light emitted from an exoplanet was first measured, paving the way for the study of exoplanetary atmospheres \citep{CharbonneauEtal2005apjTrES-1, DemingEtal2005natHD209}. In the years since, a diverse collection of worlds have been discovered, from rocky, Earth-like planets to massive gas giants that reach temperatures as hot as some stars \citep{HasegawaPudritz2013apjExoplanetPopulations, Batalha2014pnasKeplerExoplanetPopulations}. Edge-on planetary systems enable the measurement of transit (when the exoplanet passes in between the host star and the observer) and eclipse (when the exoplanet passes behind the host star as viewed by the observer) depths \citep{Kreidberg2017bookExoplanetAtmosphereMeasurements}. Transit depths measure the effective radius of the planet as a function of wavelength; variations in measured radius arise from the molecules in the atmosphere at the day-night terminator absorbing certain wavelengths of light, with more absorption corresponding to larger measured radii. Eclipse depths measure the ratio of the planet's and host star's emission as a function of wavelength. These depths provide insight into the composition and temperature structure of the planet's atmosphere.

Using measured transit or eclipse depths spanning a range of wavelengths, an atmospheric model for the planet can be determined with some uncertainty via atmospheric retrieval, an inverse modeling technique \citep{Madhusudhan2018bookAtmRetrExo}. 
Early retrieval studies performed a parametric grid search over millions of pre-calculated forward models \citep{MadhusudhanSeager2009apjAtmRetrMeth}.
This method was later improved by Bayesian techniques employing Markov chain Monte Carlo (MCMC) and other sampling techniques \citep[e.g.,][]{Skilling2004aipNestedSampling, Braak2006statcompDifferentialEvolution, Braak2008statcompSnookerDEMC} to explore a model parameter space by computing spectra for thousands to millions of atmospheric models \citep[e.g.,][]{MadhusudhanSeager2010apjThermalHotJup, LineEtal2014apjRetrievalCO, WaldmannEtal2015apjTauREx2,OreshenkoEtal2017apjlWASP12b}.
Model parameters describe the temperature--pressure profile, $T(p)$; the vertical abundance profiles for each molecule in the atmospheric model; cloud parameters; and, for the transit case, the radius of the planet.
These Bayesian techniques yield a posterior distribution which constrains the range of values that fit the data for each model parameter. For low-resolution data, some parameters may be only constrained to an upper/lower limit (or not at all) due to degeneracies among low-resolution spectra (e.g., a slightly cooler atmosphere with greater abundances of molecules will look the same as a slightly warmer atmosphere with lesser abundances). While high-resolution data allows for parameters to be more accurately determined, there is still some inherent uncertainty due to astrophysical and instrumental noise. Accurate quantification of this uncertainty informs the statistical significance of the results.

Data-driven machine learning (ML) approaches, which are able to learn complex relationships within large data sets, provide possible solutions to methods that can be computationally-expensive, such as atmospheric retrieval.
Examples using ML can be seen across the field of astrophysics from applying Bayesian linear regression to remove common-mode systematics in Kepler data \citep{RobertsEtal2013mnrasKeplerSystematicsML}, to automating the process of identifying exoplanets using deep learning \citep{ShallueVanderburg2018ajKeplerML,AnsdellEtal2018apjlKeplerML,osbornRapidTess}. 
Furthermore, in an approach similar to our own, but in a different application domain, \citet{PerreaultLevasseurEtal2017apjLensedGalaxiesBNN} used Bayesian neural networks to map distant gravitationally-lensed galaxies.

Recently, the study of exoplanetary atmospheres has been aided by ML techniques. 
\citet{Waldmann2016apjDreamingAtmospheres} makes use of deep belief networks to classify exoplanet emission spectra, importantly showing that ML approaches can identify molecular signatures in emission spectra.
The first supervised ML retrieval algorithms, {\tt HELA} \citep{MarquezNeilaEtal2018natureMLRetrieval} and {\tt ExoGAN} \citep{ZingalesWaldmann2018arxivExoGAN}, have been developed and show promising results.
{\tt HELA} uses a random forest to classify observed spectra into some planetary model (see Section \ref{sec:rf} for more details), while {\tt ExoGAN}  combines a generative adversarial network \citep[GAN,][]{Goodfellow2014nipsGAN} with a technique called semantic image inpainting \citep{yeh2017semantic} to retrieve atmospheric parameters.
These methods reduce retrieval times from hundreds of central processing unit hours to just seconds/minutes, highlighting the large reductions in computation times offered by ML.

Here, we introduce a new ML retrieval method, \texttt{plan-net}\footnote{Our code is available at \url{https://github.com/exoml/plan-net}.}, which is based on an ensemble of Bayesian neural networks, and apply it to the benchmark data set of \citet{MarquezNeilaEtal2018natureMLRetrieval}. BNNs are a good choice of model for atmospheric retrievals as they give the advantage of both providing probability distributions over their outputs and scaling to high-dimensional data. We directly compare our model with {\tt HELA} over the same data set and demonstrate how incorporating domain-specific knowledge into machine learning models can improve results and offer insights into the covariance of the atmospheric parameters.

In this paper we first introduce the data set in Section \ref{sec:data} along with the notation. We then introduce both ML models in Section \ref{sec:ml}, where we start with the random forest followed by a detailed explanation of our model. In Section \ref{sec:Res} we both display and discuss our results. Finally, in Section \ref{sec:con} we make conclusions about the implications of our results and suggest further avenues for research in this area.

\section{Data Set}\label{sec:data}
\subsection{Description}

We use the spectral data set of \citet{MarquezNeilaEtal2018natureMLRetrieval} which consists of 100,000 synthetic Hubble Space Telescope Wide Field Camera 3 (WFC3) transmission spectra of hot Jupiters. These spectra were created using the formalism detailed in \citet{HengKitzmann2017mnrasTransmissionSpectra}, which makes use of line-by-line calculations for opacities (S. Grimm, priv. comm.). This is based on five atmospheric parameters: an isothermal temperature; abundances of H$_2$O, NH$_3$, and HCN gas; and a gray cloud opacity, $\kappa_0$. Each spectrum has 13 channels with bandpasses matching those used in \citet{KreidbergEtal2015apjWASP12bWater} ($0.838$ -- $1.666$ {\textmu}m).  Each channel holds the transit depth within the corresponding bandpass. We refer the reader to their papers for more details, particularly the `Methods' section of \citet{MarquezNeilaEtal2018natureMLRetrieval}, as this is where the boundary conditions are described.

For each transit depth, we assume the same 50 parts per million uncertainty as \citet{MarquezNeilaEtal2018natureMLRetrieval}. We similarly split the data set between training ($80,000$) and testing ($20,000$). 
We reserve $10,000$ spectra from the training set to be the validation set, which is used to optimize model hyperparameters and architectures. This ensures that inferences are made on the test data only one time.
We use the same real-data test case: the WASP-12b WFC3 transit depths as analyzed by \citet{KreidbergEtal2015apjWASP12bWater}. Two sample input spectra can be seen in the Appendix, Figures \ref{fig:spectra2} and \ref{fig:spectra205}.

\subsection{Notation}

In this paper we use the following notation to describe our data set, $\mathcal{D}$. A single spectrum with $13$ channels is denoted by the vector $\mathbf{s} \in \mathcal{R}^{13}$ and $\boldsymbol{\theta} \in \mathcal{R}^5$ defines the vector of five atmospheric parameters. Furthermore, we generalize our model by referring to the dimension of $\boldsymbol{\theta}$ as $D$. The training and testing data sets are denoted by $\mathcal{D}_{\mathrm{tr}}$ and $\mathcal{D}_{\mathrm{te}}$ respectively, where the test data is given by $\mathcal{D}_{\mathrm{te}} = \{\mathbf{s}_n,\boldsymbol{\theta}_n\}_{n=1}^N$ for $N$ total input-output pairs.

\section{Machine Learning Models}\label{sec:ml}

In machine learning, the task of inferring a function from labeled data comes under the area of supervised learning. In our case, the task is a multivariate regression problem, where the objective is to model the relationship between the input-space, $\mathbf{s}$, and the output-space, $\boldsymbol{\theta}$. 
In addition to predicting the values of the outputs, it is vital that the ML model also provides an uncertainty estimation over these values. 
Astronomical observations inherently introduce uncertainty in measurements, and accurately accounting for and reporting these uncertainties is a critical part of retrieval results. 

In this section we introduce the previously-used random forest along with our \texttt{plan-net} model. In each section we explain how each model aims to solve this multivariate regression task and how they each deal with uncertainty. We highlight that the \texttt{plan-net} model is specifically designed to deal with both the uncertainty and the correlations between the outputs, whereas the the random forest does not differ from those used in other multivariate regression tasks.
\newpage
\subsection{Random Forest}
\label{sec:rf}

Here, we briefly summarize the random forest regression model used in \citet{MarquezNeilaEtal2018natureMLRetrieval}, where the details of the model are available at \url{https://github.com/exoclime/HELA}. The core of their model comes from the \texttt{\small ensemble.RandomForestRegressor} method in \texttt{\small sklearn} \citep{scikit-learn}. 
A random forest (RF) consists of multiple decision trees (or regression trees, for the case of continuous data), whereby each tree makes a prediction given an input (see \cite{criminisi2012decision}). \citet{MarquezNeilaEtal2018natureMLRetrieval} showed that no more than $1,000$ regression trees were required, which led to choosing that number for the model. They set the number of nodes in each tree via a variance threshold of $0.01$. This is a metric that is related to the proportion of the remaining training data that is split at the current node.

To produce the posterior plots, as shown in Figure \ref{fig:rf_wasp}, each prediction from a tree corresponds to a sample from an empirical distribution. The $1000$ samples therefore correspond to the density estimation of the atmospheric parameters.

\subsection{Bayesian Neural Networks}
Our model is built from Bayesian neural networks, which inherit their structure from neural networks. Although we provide details of both techniques in the following section, we highlight their strong relationship with multivariate linear regression, where the objective is to learn a matrix of weights $\mathbf{W}$ that map an input $\mathbf{s}$ to an output $\boldsymbol{\theta}$. Fully connected deep neural networks extend upon this by combining layers of linear regression with non-linear functions to result in a more powerful function-approximating capability, despite still operating on the same supervised learning task as a linear regression model. 

\subsubsection{A Summary}\label{sec:BNN_sum}

Bayesian neural networks (BNNs) offer the powerful function-approximating capability of deep neural networks with the additional advantage of being able to provide distributions over their outputs \citep{mackay1992practical, neal1995bayesian}.
Therefore, these characteristics are well-suited to the task of atmospheric retrieval. To enable BNNs to scale to large architectures we employ the Monte Carlo dropout approximation to BNNs \citep{gal2016dropout}. This is a stochastic variational inference approach \citep{hoffman2013stochastic} that allows BNN inference to be performed for both large architectures and large data sets. The alternative approach would be to implement a form of MCMC such as Hamiltonian Monte Carlo \citep[HMC,][]{neal1995bayesian} to perform inference. Although HMC has been shown to be successful at small scale, it currently cannot be scaled in the same way as stochastic variational inference approaches.

Deep neural networks consist of a hierarchy of layers, where each layer applies a non-linear weighted transformation of its input. We define each layer $l$, to have its own matrix of weights $\mathbf{W}_l$ and biases $\mathbf{b}_l$. If $\mathbf{h}(\cdot)$ is a non-linear function then we can define a fully connected dense neural network with $L$ layers and input $\mathbf{s}$ as: $$\mathbf{f}^{\boldsymbol{\omega}}(\mathbf{s}) =\mathbf{W}_L\mathbf{h}\left(\dots \mathbf{h}(\mathbf{W}_0 \mathbf{s} + \mathbf{b}_l)\dots\right) + \mathbf{b}_L,$$
where $\boldsymbol{\omega}= \{\mathbf{W}_l,\mathbf{b}_l\}_{l=1}^L$ and refers to all the network weights.
A BNN takes this formulation and adds a prior $p(\boldsymbol{\omega})$ over the weights, often taking the form of a multivariate normal distribution. Bayesian inference in BNNs requires computing an intractable integral to infer $p(\boldsymbol{\omega}\mid \mathcal{D}_{\mathrm{tr}})$. The Monte Carlo dropout approximation provides a (variational) approximation to this distribution and comes under the wider area of variational inference \citep{jordan1998introduction}. Practical implementation of MC dropout requires drawing dropout masks \citep{srivastava2014dropout} from Bernoulli-distributed random variables to set a certain proportion of weights to zero. Applying this during the training of the network acts as a regularizer to prevent overfitting. Dropping these weights whilst making predictions at test-time results in the test-time approximation for predictions over the outputs. For a given input $\mathbf{s}_n$, we can sample the network $T$ times to result in an empirical distribution $p(\boldsymbol{\theta}\mid \mathbf{s}_n,\mathcal{D}_{\mathrm{tr}})$.

Determining the proportion of weights to be dropped in each layer $p_l$ often requires tuning over a validation set. However, we use concrete dropout layers to automatically optimize for these values in the training process \citep{gal2017concrete}.

\subsubsection{The Model}

Our model, \texttt{plan-net}, shown in Figure \ref{fig:model}, is a deep neural network with four dense concrete dropout layers \citep{gal2017concrete}. The model is implemented in \texttt{Keras} \citep{chollet2015keras} with a \texttt{TensorFlow} backend \citep{abadi2016tensorflow}. Each layer consists of 1024 units, and we use a batch size of 512. For training the model, we use the Adam optimization algorithm \citep{kingma2014adam}. For deciding on the architecture, we implemented a grid search over the number of layers and the number of units per layer.

Our task is to accurately predict the atmospheric parameters and provide posterior\footnote{In the machine learning literature, the output distribution would normally be called the predictive distribution as we are inferring the posterior over the weights of the network and then working with this posterior to infer a predictive distribution. However, to remain consistent with the exoplanet literature, we avoid that here.} distributions over their values.
These parameters are expected to covary and we directly use this domain knowledge to design our model, such that we can represent the atmospheric parameters to be jointly distributed by a multivariate normal distribution.
Therefore we design the output of the BNN to consist of the a lower triangular matrix \(\mathbf{L}\) of dimensions $D\times D$ and a mean vector \(\boldsymbol{\mu}\) of dimension $D$. We can then represent the precision matrix of a multivariate normal via its Cholesky decomposition \(\boldsymbol{\Lambda} = \mathbf{L}\mathbf{L}^{\top} \).

Figure \ref{fig:model} demonstrates the atmospheric retrieval process after the model is trained. We implement $T$ forward passes through the network for a given observed spectrum $\mathbf{s}_n$, resulting in the samples $\{\boldsymbol{\mu}(\mathbf{s}_n)_t, \mathbf{L}(\mathbf{s}_n)_t  \}_{t=1}^T$. 
In the next step, we take the mean over these network samples to give the expected $\mathbf{L}(\mathbf{s}_n)$ and $\boldsymbol{\mu}(\mathbf{s}_n)$ for a given spectrum:
\begin{equation}
    \mathbf{L}(\mathbf{s}_n) = \frac{1}{T}\sum_{t=1}^T \mathbf{L}(\mathbf{s}_n)_t,
\end{equation}
\begin{equation}
    \boldsymbol{\mu}(\mathbf{s}_n) = \frac{1}{T}\sum_{t=1}^T \boldsymbol{\mu}(\mathbf{s}_n)_t.
\end{equation} 
The final step is to sample from the multivariate normal distribution,
\begin{equation}\label{eq:mult_norm}
\boldsymbol{\theta} \sim\mathcal{N}\left(\boldsymbol{\mu}(\mathbf{s}_n)
,(\mathbf{L}(\mathbf{s}_n)\mathbf{L}^{\top}(\mathbf{s}_n))^{-1}\right)
\end{equation}
to retrieve samples from the inferred atmospheric parameters, where this distribution is parameterized by the expectation BNN output.

\subsubsection{Training}

\begin{figure}
    \centering
        \includegraphics[width=\columnwidth]{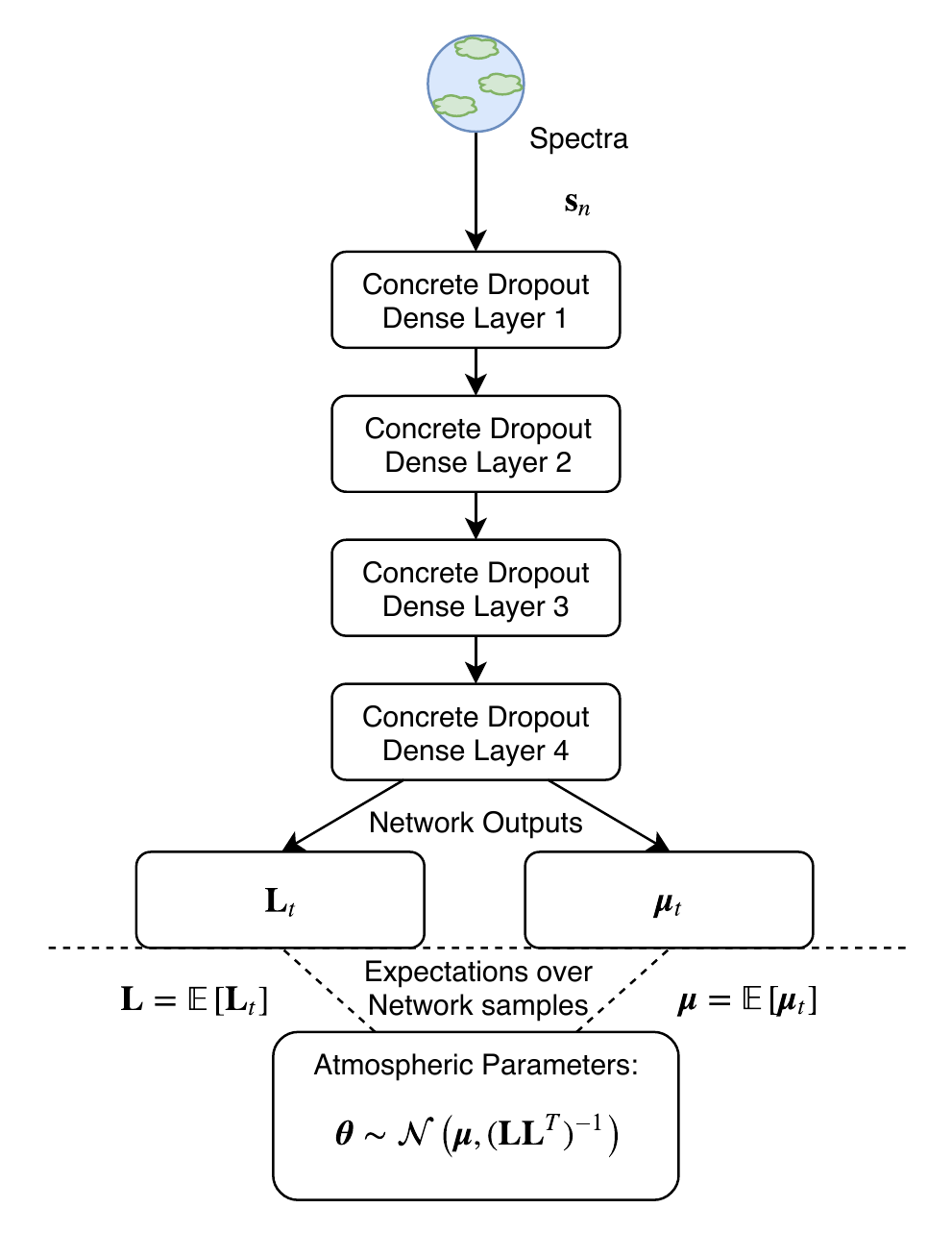}
    \caption{\texttt{plan-net} model procedure at test time for a given spectrum $\mathbf{S}_n$. $T$ samples are taken from the BNN and the expectations over the lower triangular matrix and the mean are then used to parameterize the multivariate normal distribution. $\boldsymbol{\theta}$ can then be drawn from this distribution to retrieve the atmospheric parameters. Each concrete dropout layer consists of 1024 units. 
    }\label{fig:model}
     \vspace{-.01in}
\end{figure}

In order to train this model, we must design a loss that ensures the network learns the correlations between the atmospheric parameters. In order to estimate the covariance, our loss is the negative log-likelihood of the multivariate normal, as defined by $\boldsymbol{\mu}$ and $\mathbf{L}$. The loss,
\begin{equation}\label{eq:loss}
\mathcal{L}(\boldsymbol{\omega},\boldsymbol{\mu},\mathbf{L}) = -2\sum_{d=1}^D \log (l_{dd}) + (\mathbf{y} - \boldsymbol{\mu})^{\top}\mathbf{L}\mathbf{L}^{\top}(\mathbf{y} - \boldsymbol{\mu}),
\end{equation}
is defined to be implicitly dependent on the network weights $\boldsymbol{\omega}$ through the lower triangular matrix $\mathbf{L}$ and the inferred mean $\boldsymbol{\mu}$ (see Figure \ref{fig:model}). As also mentioned in \citet{dorta2018structured}, we must be careful to ensure that the diagonal elements, $l_{ii}$, of $\mathbf{L}$ are positive such that $\boldsymbol{\Lambda}$ is positive-definite; we therefore take the exponential of the diagonal terms to ensure this. In comparison to previous loss functions that have been used for BNNs, such as the squared loss and the heteroscedastic squared loss (see \citet[Chapter~4]{gal2016uncertainty}), our new loss in Equation \eqref{eq:loss} is able to model correlations between atmospheric parameters. These inferred correlations lead to better uncertainty estimates for the retrieved atmospheric parameters than the previous losses.

In addition to using the Adam optimizer, we employ early stopping, with a patience of 30 epochs, according to the validation loss. Furthermore, we use model checkpointing to save the model that has the best performance on the validation set.

\subsection{Ensemble}

It has been shown that an ensemble of neural networks can offer more accurate estimations of the predictive uncertainty than a single network \citep{lakshminarayanan2017simple, gal2018sufficient}. The additional benefit is that an ensemble is more robust to changes in weight initialization and the path taken during stochastic optimization.

In this paper we use an ensemble of five \texttt{plan-net} models and provide comparison to a single model. Five models were chosen due to the empirical performance in Table \ref{tab:R2}, as larger ensembles result in increasingly marginal improvements.

The challenge in using an ensemble is in how the outputs from the individual models are combined. In our case, each output is the mean and covariance of a multivariate normal distribution. Therefore in combining these distributions together, we can treat the overall output from the ensemble as a Gaussian mixture model, whereby the each component weight corresponds to $^1/_M$, where $M$ is the number of models in the ensemble.

To calculate the expectation of this mixture model, $\boldsymbol{\mu}_\mathrm{ens}$, we take the average of the individual component means such that
$$\boldsymbol{\mu}_{\mathrm{ens}}  = \frac{1}{M}\sum_{m=1}^M\boldsymbol{\mu}_m.$$
The variance of the mixture model $\boldsymbol{\Sigma}_{\mathrm{ens}}$ can be calculated by employing the law of total variance:
$$\boldsymbol{\Sigma}_{\mathrm{ens}} = \frac{1}{M}\sum_{m=1}^M \left(\boldsymbol{\mu}_\mathrm{m} - \boldsymbol{\mu}_\mathrm{ens} \right)^2 + \frac{1}{M} \sum_{m=1}^M \boldsymbol{\Sigma}_{m},$$
where the inferred covariance matrix of a single model is given by \(\boldsymbol{\Sigma}_{m} = \boldsymbol{\Lambda}_m^{-1} = (\mathbf{L}_m\mathbf{L}_m^{\top})^{-1} \).
This combines the variance in the component means with the expectation of the variance of the individual models, thus taking into account how unsure each model is and how far each model's mean lies from the ensemble mean. 
Therefore the atmospheric parameters retrieved via the ensemble $\boldsymbol{\theta}_{\mathrm{ens}}$ are distributed according to $\boldsymbol{\theta}_{\mathrm{ens}} \sim \mathcal{N}\left(\boldsymbol{\mu}_{\mathrm{ens}}, \boldsymbol{\Sigma}_{\mathrm{ens}} \right)$.

\begin{figure}[t]
\captionsetup[subfloat]{farskip=2pt,captionskip=1pt}
    \centering
    \subfloat[Random Forest]{%
 \includegraphics[width=.34\textwidth]{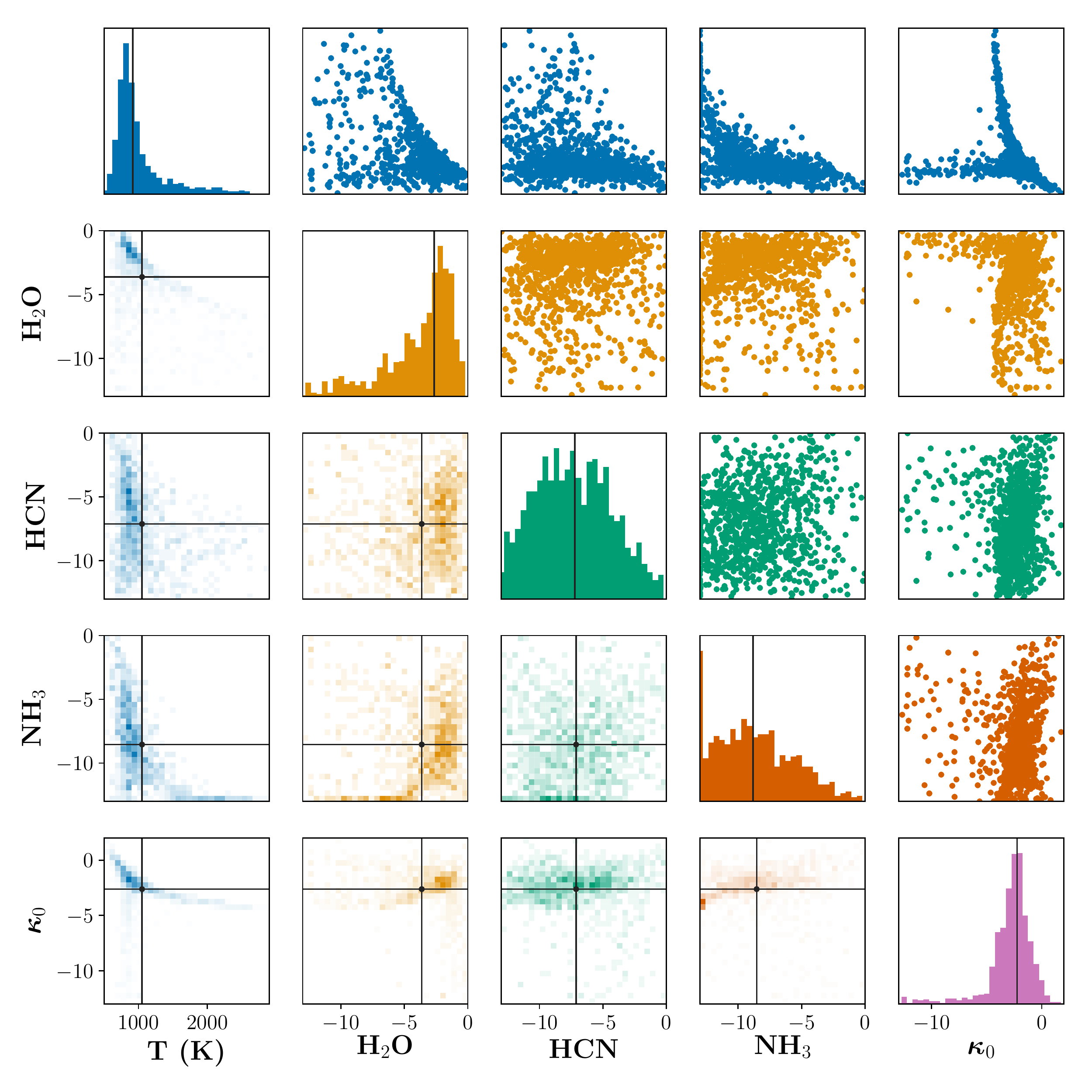}%
  \label{fig:rf_wasp}%
  }\qquad
  \subfloat[\texttt{plan-net}]{%
  \includegraphics[width=.34\textwidth]{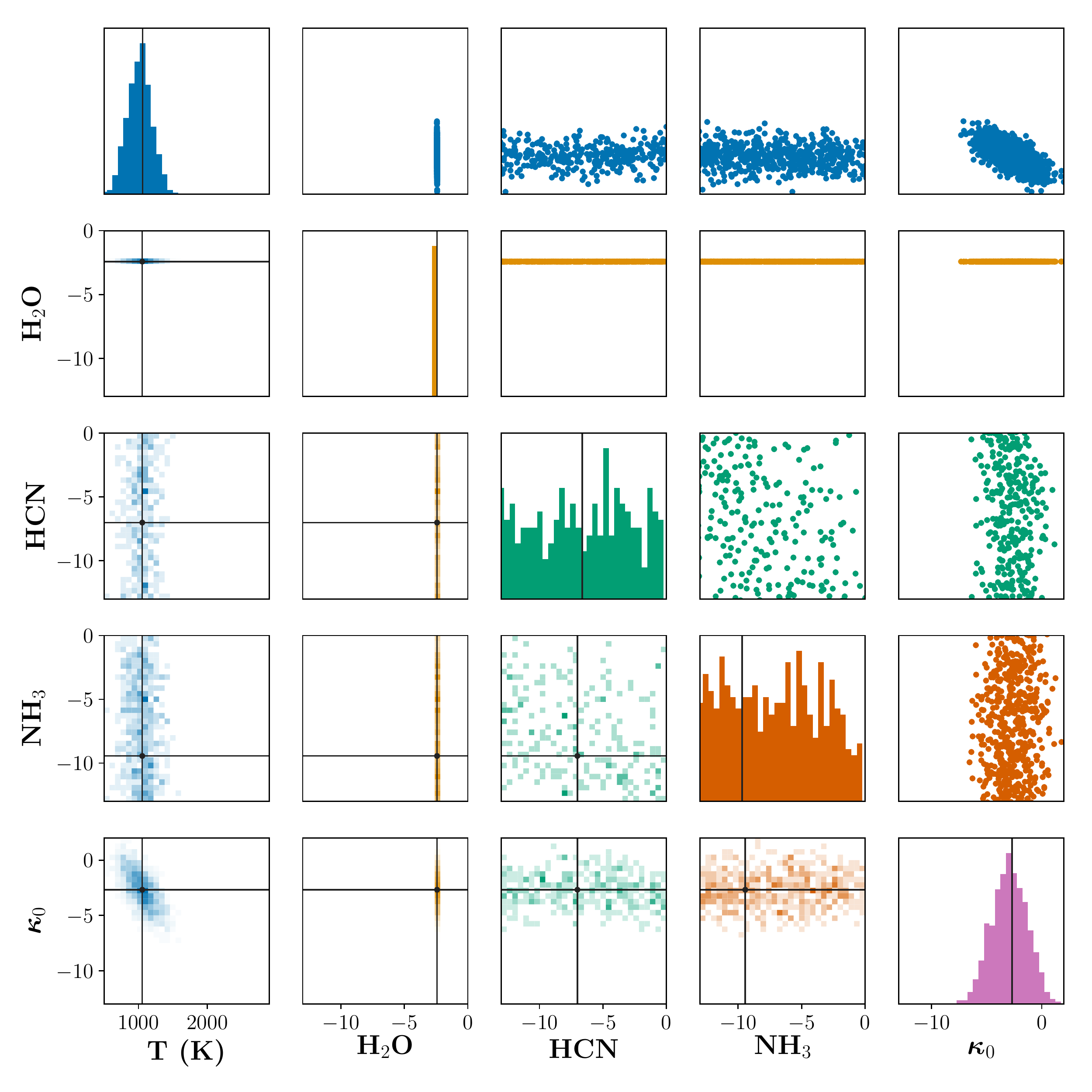}%
  \label{fig:bnn_wasp}
  }\qquad 
  \subfloat[\texttt{plan-net} Ensemble]{%
  \includegraphics[width=.34\textwidth]{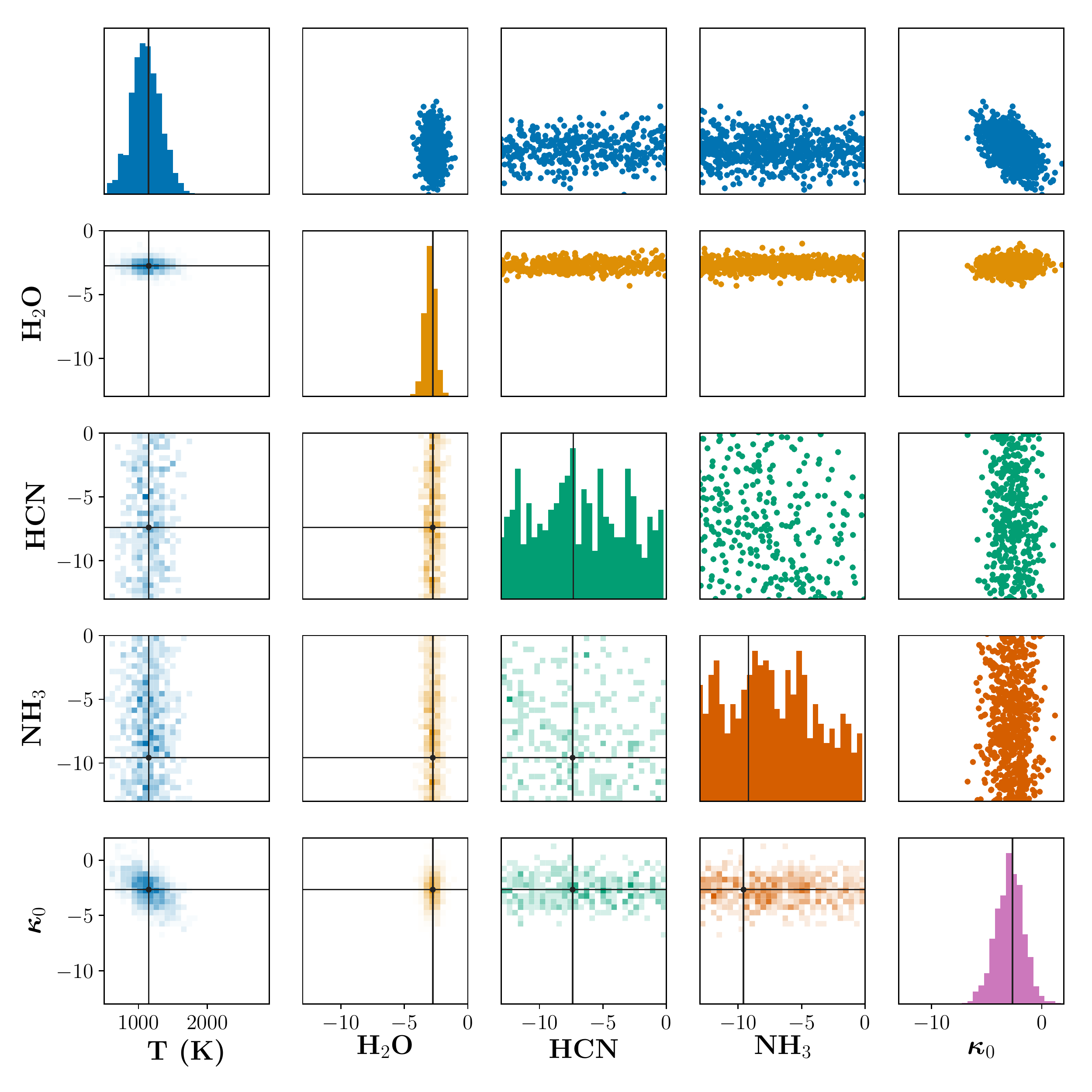}%
  \label{fig:bnn_ens_wasp}
  }
    \caption{Retrieval analysis of the WFC3 transmission spectrum of WASP-12b, where we compare the random forest with both a single \texttt{plan-net} and a \texttt{plan-net} ensemble. The black cross denotes the mean over the samples, where we report the results in Table \ref{tab:wasp}. We note consistent results across all models, and highlight the broader posteriors of the ensemble when comparing to the single \texttt{plan-net}.}\label{fig:wasp}
\end{figure}

\section{Results and Discussion}\label{sec:Res}
Table \ref{tab:R2} displays a comparison of $R^2$ values across the models, where $R^2$ corresponds to the coefficient of determination
\begin{equation}
R^2 = 1 - \frac{\sum_{n=1}^N \sum_{d=1}^D\left(\theta_{n}^{(d)} - \mu_{\mathrm{ens}}^{(d)}(\mathbf{s}_n)\right)}{\sum_{n=1}^N \sum_{d=1}^D(\theta_{n}^{(d)} - \tilde{\theta}^{(d)})}
\end{equation}
as defined in the \texttt{\small sklearn.metrics} Python package, where the summation is over both the size of the data set $N$ and the output dimension $D$. $\tilde{\theta}^{(d)}$ is the data mean for each atmospheric parameter and the prediction for each data point is given by $\mu_{\mathrm{ens}}^{(d)}(\mathbf{s}_n)$. This can be viewed as a ratio between the residuals for the model prediction and the total sum of squares.
Values close to $1.0$ are desirable as they are related to the correlation coefficient between the predicted and true atmospheric parameters. 

Therefore, the results in Table \ref{tab:R2} show that both of our models, the ensemble and the individual \texttt{plan-net} model, outperform the random forest. Furthermore, we note the slight performance boost that is gained from the ensemble. In order to show that the results are reproducible, we list both our implementation of the random forest and their reported results, which closely agree.

In addition to reporting the $R^2$ values, Table \ref{tab:cov} contains the average covariance matrix over the test data. This table shows the average inferred correlations, where the diagonal corresponds to the variance in each atmospheric parameter and the off-diagonals indicate correlations between the parameters. As this is the average correlation matrix for all $20,000$ test planets, not too many conclusions can be drawn from this matrix. However, we note the average negative correlation that appears between $T(\mathrm{K})$ and $\kappa_0$ as well as $T(\mathrm{K})$ and H$_2$O. This is consistent with intuition due to the known degeneracies in the data. More specifically, as the observed spectral features are caused by the temperature--pressure profile and the molecular abundances, increasing either whilst keeping all other parameters constant leads to stronger spectral features. Consequently, a simultaneous increase in temperature and a decrease in molecular abundances (or vice versa) could lead to the same observed spectrum. Finally, an increase in cloud opacity decreases the intensity of the observed spectral features and could therefore look similar to a decrease in temperature, hence the degeneracy and the expected negative correlation between $T(\mathrm{K})$ and $\kappa_0$.

\begin{table}[t]
\caption{Table reports $R^2$ values for each atmospheric parameter. Values near $1$ indicate high correlation between model prediction and the known atmospheric parameters. \texttt{plan-net} achieves a higher overall mean $R^2$ as well as being higher for each individual parameter. Bold indicates the best $R^2$ value for each parameter.
}\vskip -.1in
\label{tab:R2}
\setlength{\tabcolsep}{3.3pt}
\begin{center}
\begin{sc}
\begin{tiny}
\begin{tabular}{ l  c c c c c c }
 \toprule
& $T(\mathrm{K})$ & $\mathrm{log} X_{\mathrm{H}_2\mathrm{O}}$ &$\mathrm{log} X_{\mathrm{HCN}}$ &$\mathrm{log} X_{\mathrm{NH}_3}$ &$ \kappa_0$ & Mean\\
\hline
\texttt{plan-net} $R^2$ &$\mathbf{0.770}$	&	$0.623$  & $0.487$ & $0.721$ & $0.750$&$\mathbf{0.673}$\\
 Ens. 5 \texttt{plan-net} $R^2$ &$\mathbf{0.770}$	&	$\mathbf{0.629}$  & $\mathbf{0.491}$ & $\mathbf{0.723}$ & $\mathbf{0.751}$&$\mathbf{0.673}$\\
Our Ran. Forest $R^2$ &$0.746$   &$0.608$   &$0.466$ &$0.700$ &$0.736$&$0.651$\\
Ran. Forest\footnote{Reported from \citet{MarquezNeilaEtal2018natureMLRetrieval}.} $R^2$ &$0.746$   &$0.608$   &$0.467$ &$0.700$ &$0.737$&$0.652$\\

\end{tabular}
\end{tiny}
\end{sc}
\end{center}
\end{table}

\begin{table}[t]
\caption{Mean inferred normalized correlation matrix, $(\mathbf{L}\mathbf{L}^{\top})^{-1} $, across all test set atmospheric retrievals. The diagonal values are the mean marginalized variances for each parameter. The off-diagonals indicate correlations between these parameters; note the expected negative correlation between $T(\mathrm{K})$ and $\kappa_0$ as well as $T(\mathrm{K})$ and H$_2$O.}\vskip -.1in
\label{tab:cov}
\setlength{\tabcolsep}{3.3pt}
\begin{center}
\begin{sc}
\begin{scriptsize}
\noindent\begin{tabular}{c*{5}{|m{1.cm}}|}
 \multicolumn{1}{c}{} & \multicolumn{1}{c}{$T(\mathrm{K})$} & \multicolumn{1}{c}{$\mathrm{H}_2\mathrm{O}$} 
  & \multicolumn{1}{c}{$\mathrm{HCN}$} & \multicolumn{1}{c}{$\mathrm{NH}_3$} & \multicolumn{1}{c}{$\kappa_0$}  \\ \hhline{~*5{|-}|}
 $T(\mathrm{K})$ & $5.43$ & $-7.50$ & $-3.30$& $-4.88$& $-0.498$  \\ \hhline{~*5{|-}|}
 $\mathrm{H}_2\mathrm{O}$ & $-7.50$ & $32.6$ & $0.454$ & $4.47$ & $0.566$ \\ \hhline{~*5{|-}|} 
 $\mathrm{HCN}$ & $-3.30$ & $0.454$ & $56.7$ & $1.37$ & $1.95$\\ \hhline{~*5{|-}|}
 $\mathrm{NH}_3$ & $-4.88$ & $4.47$ & $1.37$ & $12.1$ & $0.965$ \\ \hhline{~*5{|-}|}
 $\kappa_0$ & $-0.498$ & $0.566$ & $1.95$ & $0.965$ & $3.74$  \\ \hhline{~*5{|-}|}
\end{tabular}\vskip -.1in
\end{scriptsize}
\end{sc}
\end{center}
\end{table}

\begin{table*}[t]
\caption{Retrieved atmospheric parameters for WASP-12b. All retrievals are consistent, with our ensemble \texttt{plan-net} model achieving closer agreement with the temperature and H$_2$O abundance retrieved by \citet{KreidbergEtal2015apjWASP12bWater}. We note that \citet{KreidbergEtal2015apjWASP12bWater} did not retrieve for $\log X _{\mathrm{HCN}}$ and $\log X _{\mathrm{NH}_3}$. They also used a different cloud parameterization that makes $\kappa_0$ not applicable to their model. Errors are reported for one standard deviation, where we report the median and equivalent asymmetric posterior percentiles for the random forest and for \citet{KreidbergEtal2015apjWASP12bWater}. }
\label{tab:wasp}
\setlength{\tabcolsep}{3.3pt}
\begin{center}
\begin{sc}
\begin{scriptsize}
\begin{tabular}{ l  c c c c c }
 \toprule
& $T(\mathrm{K})$ & $\log X _{\mathrm{H}_2\mathrm{O}}$ &$\log X _{\mathrm{HCN}}$ &$\log X _{\mathrm{NH}_3}$ &$ \kappa_0$ \\
\hline
\citet{KreidbergEtal2015apjWASP12bWater}  &$1371^{+466}_{-343}$	& $-2.7^{+1.0}_{-1.1}$	  & - & -&- \\
\citet{MarquezNeilaEtal2018natureMLRetrieval} nested sampling & $1105^{+545}_{-287}$ & $-3.0^{+2.0}_{-1.9}$ & $-8.5^{+3.8}_{-2.9}$ & $-8.4^{+3.1}_{-2.9}$ & $-2.8\pm 0.9$\\
Our Rand. Forest &$937^{+410}_{-146}$   &$-2.835^{+1.51}_{-3.37}$   &$-7.484^{+3.43}_{-2.89}$ &$-9.202^{+4.12}_{-2.74}$ &$-2.281^{+1.09}_{-1.57}$\\
  Ens. 5 \texttt{plan-net} &$1142\pm 412$	&	$-2.781\pm 0.429$  & $-8.210\pm 12.7$ & $-9.605\pm 6.7$ & $-2.601 \pm 1.23$\\

\end{tabular}
\end{scriptsize}
\end{sc}
\end{center}

\end{table*}

By designing our model to learn these correlations, we are able to interpret the results in a way that is not always available when using deep learning models. Specifically, we identify cases where both our model and the random forest approach do not recover the true values, but where our model includes the true values in its wider posterior distributions. Figure \ref{fig:plan_1} shows a case where the random forest infers narrow (highly confident) posterior distributions that fall far from the true values, whereas our \texttt{plan-net} ensemble model is (appropriately) less confident, leading to posterior distributions that cover the true values for the atmospheric parameters (shown by the red stars).

Given the performance over the synthetic test data set, we further test our models on the WFC3 transmission spectrum of WASP-12b. Figure \ref{fig:wasp} shows the posterior plots for the random forest, the single \texttt{plan-net} model and the \texttt{plan-net} ensemble.
In the case of WASP-12b, both \texttt{plan-net}-based models find marginalized posteriors similar to the random forest for the cloud opacity ($\kappa_0$) and the abundances of HCN and NH$_3$. For temperature, both \texttt{plan-net}-based models have a distribution that is consistent with the retrieval performed in \citet{KreidbergEtal2015apjWASP12bWater}, while the random forest favors cooler temperatures. All models favor low ($\leq 10^{-7}$) abundances for HCN and NH$_3$, indicating a non-detection of these molecules. The H$_2$O abundance predicted by the individual \texttt{plan-net} model and the ensemble are more tightly constrained than the results of \citet{MarquezNeilaEtal2018natureMLRetrieval} or \citet{KreidbergEtal2015apjWASP12bWater}; see Table \ref{tab:wasp} for numerical comparisons\footnote{\cite{MarquezNeilaEtal2018natureMLRetrieval} utilize a constant-opacity cloud parameterization, while \citet{KreidbergEtal2015apjWASP12bWater} use a cloud and haze model that assumes an opaque gray cloud deck, which introduces degeneracies between the cloud and haze parameters. Consequently, a direct comparison between the two models cannot be made in Table \ref{tab:wasp}.
}. 
\citet{FisherHeng2018mnrasWFC3TransmissionSpectraDegeneracy} found that, in general, WFC3 transmission spectra are adequately explained by an isothermal atmosphere (in the regions probed by transit observation), gray clouds, and H$_2$O only. Based on our ensemble's confidence in H$_2$O abundance (and lack of confidence in HCN and NH$_3$ abundances), it is likely that the model similarly learned this.

\subsection{Limitations}

We highlight that employing variational approximate inference in BNNs is known to have problems, particularly underestimating uncertainty \citep{blei2017variational}. Unlike the RF, our ensemble BNN model favors large uncertainties when the data cannot constrain a parameter, as shown in Figures \ref{fig:plan_1} and \ref{fig:205}. Though an ensemble of models helps to improve the uncertainty estimation, we emphasize that accurate uncertainty estimation requires using MCMC, nested sampling, or another Bayesian sampling algorithm proven to obtain accurate posterior distributions and therefore uncertainty estimations \citep[e.g., ][]{Braak2008statcompSnookerDEMC}.

Nevertheless, BNNs are presently an important tool for retrievals. They provide a reasonable estimation of parameters orders of magnitude faster than traditional methods that require hundreds of hours of CPU time, helping to constrain parameter spaces. As an example, a single \texttt{plan-net} prediction over a test planet takes $29.4$ ms, when $T = 30$ samples, and an ensemble of five takes $1.5$ s if they are run sequentially.\footnote{Hardware: Ubuntu 18.04, 32GB memory, CPU: Intel Core i7-8700K, GPU: TITAN Xp}  

As long as the data set used to train the model contains all relevant molecules, BNNs can inform which molecules should be considered in a traditional retrieval analysis based on retrieved abundances and their uncertainties. A single \texttt{plan-net} must be trained once for a certain class of planets, e.g., WFC3 transmission spectra of hot Jupiters. Once the model has been trained, all inferences with that model are fast and repeatable, for the class of planets represented in the training set. Therefore, although training the model can be computationally expensive, this only needs to be done once. In our example, each \texttt{plan-net} model takes $20$ minutes to train over the WFC3 transmission spectra. Thus, despite the limitations of BNNs, their results are valuable and help save compute time spent on retrieval analyses.

Our approach is a generalizable technique that is not limited to any specific type of planet. In addition, important parameters such as the radius of the planet should be included in future models, as in this paper we make use of the data set of \citet{MarquezNeilaEtal2018natureMLRetrieval} which does not include it in the parameter space. Therefore the challenge in using BNNs comes from ensuring that the data set contains both the parameter space and planet types of interest.
\newpage
\section{Conclusions}\label{sec:con}
In this paper, we have demonstrated how domain-knowledge can be used to design a machine learning model that both outperforms the previous approach and provides inferred correlations between its outputs. 
Furthermore, we have introduced a novel likelihood function for BNNs which captures correlations between output dimensions. 
This extends on the diagonal Gaussian likelihood often used in the literature that does not capture these correlations. We highlight that this is extremely easy to do with BNNs and stochastic approximate inference, when comparing to traditional ML techniques (e.g. Gaussian processes), where it would involve many more approximations.

Using the data set of \citet{MarquezNeilaEtal2018natureMLRetrieval}, we independently reproduced the results of their RF. For the first time, we have shown that ML retrieval results are reproducible and consistent across implementations.

In addition to comparing our approach to the random forest using $20,000$ test planet models, we also analyzed the inferred posteriors for WASP-12b, where we take the results of \citet{KreidbergEtal2015apjWASP12bWater} to be the ground truth. Our ensemble of five \texttt{plan-net} models gives results consistent with the RF of \citet{MarquezNeilaEtal2018natureMLRetrieval} and achieved distributions for H$_2$O abundance and temperature that agree more closely with \citet{KreidbergEtal2015apjWASP12bWater} and the nested sampling retrieval of \citet{MarquezNeilaEtal2018natureMLRetrieval} than the RF. The low ($< 10^{-7}$) retrieved abundances and large uncertainties of HCN and NH$_3$ indicate a non-detection of these molecules.

Furthermore, we have found that an ensemble of BNNs provides posterior distributions that better represent those of traditional Bayesian atmospheric retrieval methods, compared to both a single BNN model and the RF model. A single \texttt{plan-net} model can underestimate the size of the posterior distributions due to overconfidence in their predictions, while the RF can be overconfident in a wrong answer.

We have presented the first study that employs BNNs for atmospheric retrievals, setting the foundation for further research in this area. As the data available for atmospheric retrievals expands, it will become increasingly important to combine domain-knowledge with machine learning models. It is equally important that it remains possible to interpret the outputs of these models so that inferences can be physically understood. Our method easily scales to higher dimensionality; in future work, we will expand our model to higher resolution spectra and a larger number of atmospheric parameters.

\acknowledgements

\noindent We thank Chloe Fisher for making the data set from \citet{MarquezNeilaEtal2018natureMLRetrieval} publicly available on GitHub upon request. Adam D. Cobb is sponsored by the AIMS CDT (\url{http: //aims.robots.ox.ac.uk}) and the EPSRC (\url{https: //www.epsrc.ac.uk}). Frank Soboczenski gratefully acknowledges the support of NVIDIA Corporation with the donation of the Titan Xp GPU used for this research (GPU No 900-1G611-2530-000). A.G. Baydin is funded by Lawrence Berkeley National Lab and EPSRC/MURI grant EP/N019474/1. We thank NASA FDL (\url{http: //www.frontierdevelopmentlab.org/}) and SETI (\url{https://www.seti.org}) for making this collaboration possible. 

\bibliography{INARA-W12}

\begin{appendix}
\twocolumngrid
\label{sec:appendix}

\begin{figure}[h]
    \centering
    \subfloat[Random Forest]{%
 \includegraphics[width=.35\textwidth]{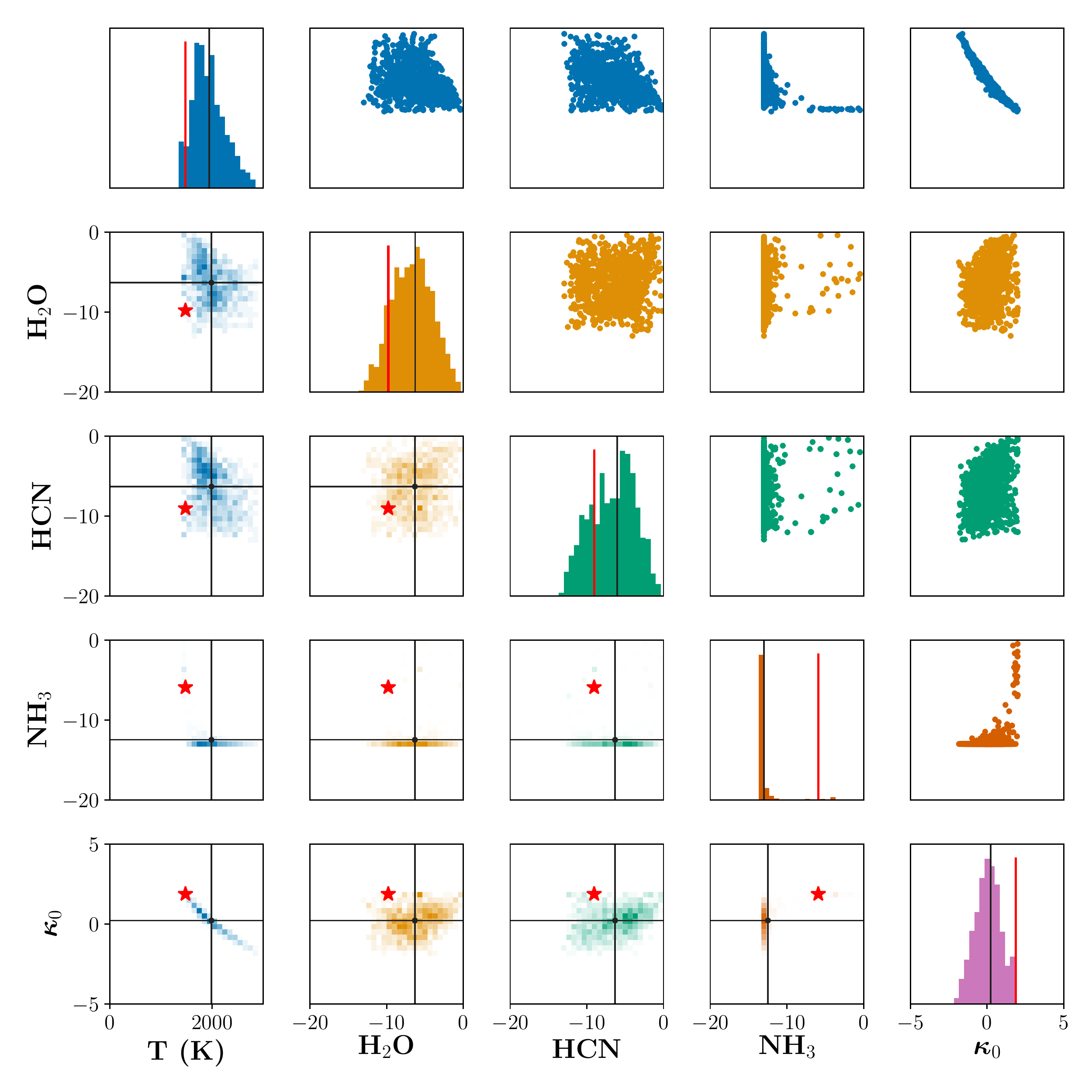}%
  \label{fig:rf_2}%
  }\qquad
  \subfloat[\texttt{plan-net} Ensemble]{%
  \includegraphics[width=.35\textwidth]{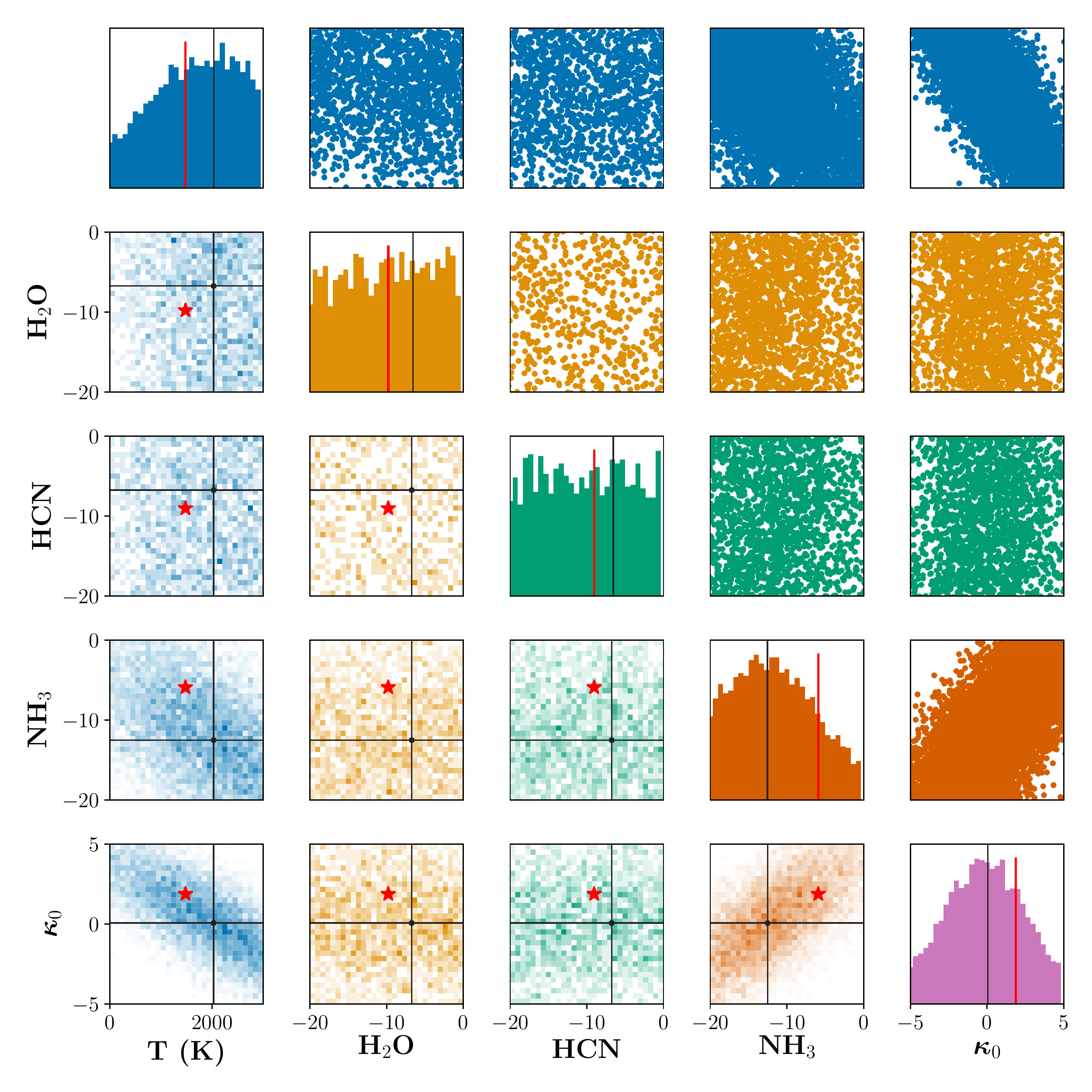}%
  \label{fig:bnn_ens2}
  }\qquad
  \subfloat[Test Spectrum]{%
  \includegraphics[width=.35\textwidth]{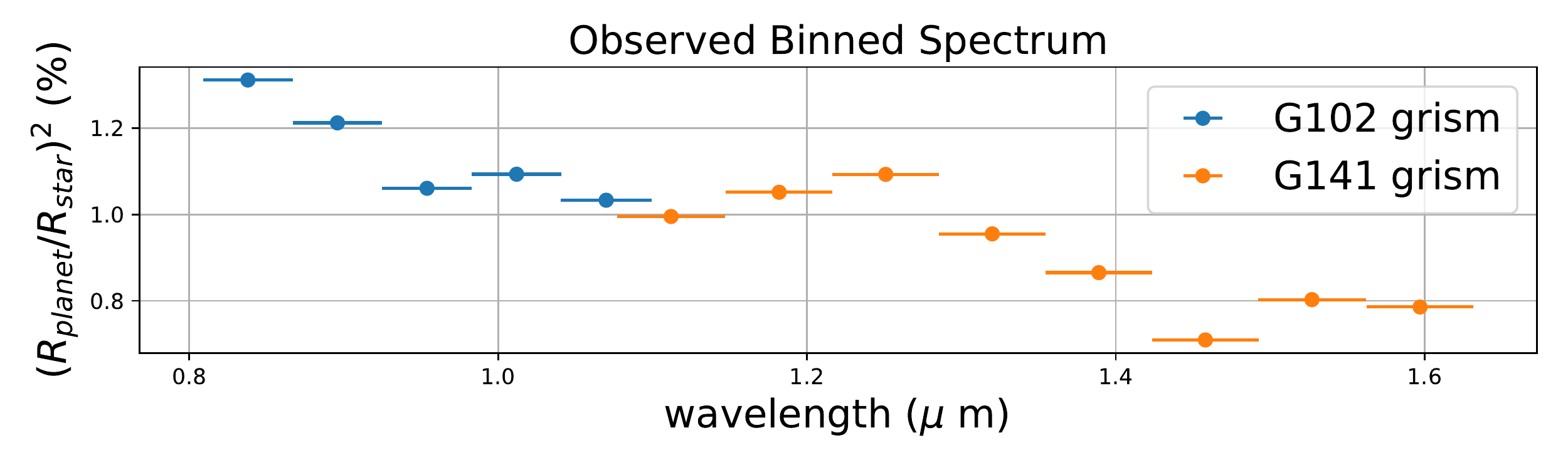}%
  \label{fig:spectra2}
  }\qquad
    \caption{Test planet 1: an example taken from the test set,
    where the random forest is overconfident and far from the true parameter values, denoted by the red star. In comparison, the \texttt{plan-net} ensemble demonstrates its uncertainty in its  predicted values by inferring broader posterior distributions that cover the true parameters. Figure \ref{fig:spectra2} gives the observed input spectrum, where the binning is given by Table 3 in \cite{KreidbergEtal2015apjWASP12bWater}. Each spectral coverage of the wavelengths is given by two grisms indicated in the legend.}\label{fig:plan_1}
\end{figure}

\begin{figure}[h]
    \centering
    \subfloat[Random Forest]{
\includegraphics[width=.35\textwidth]{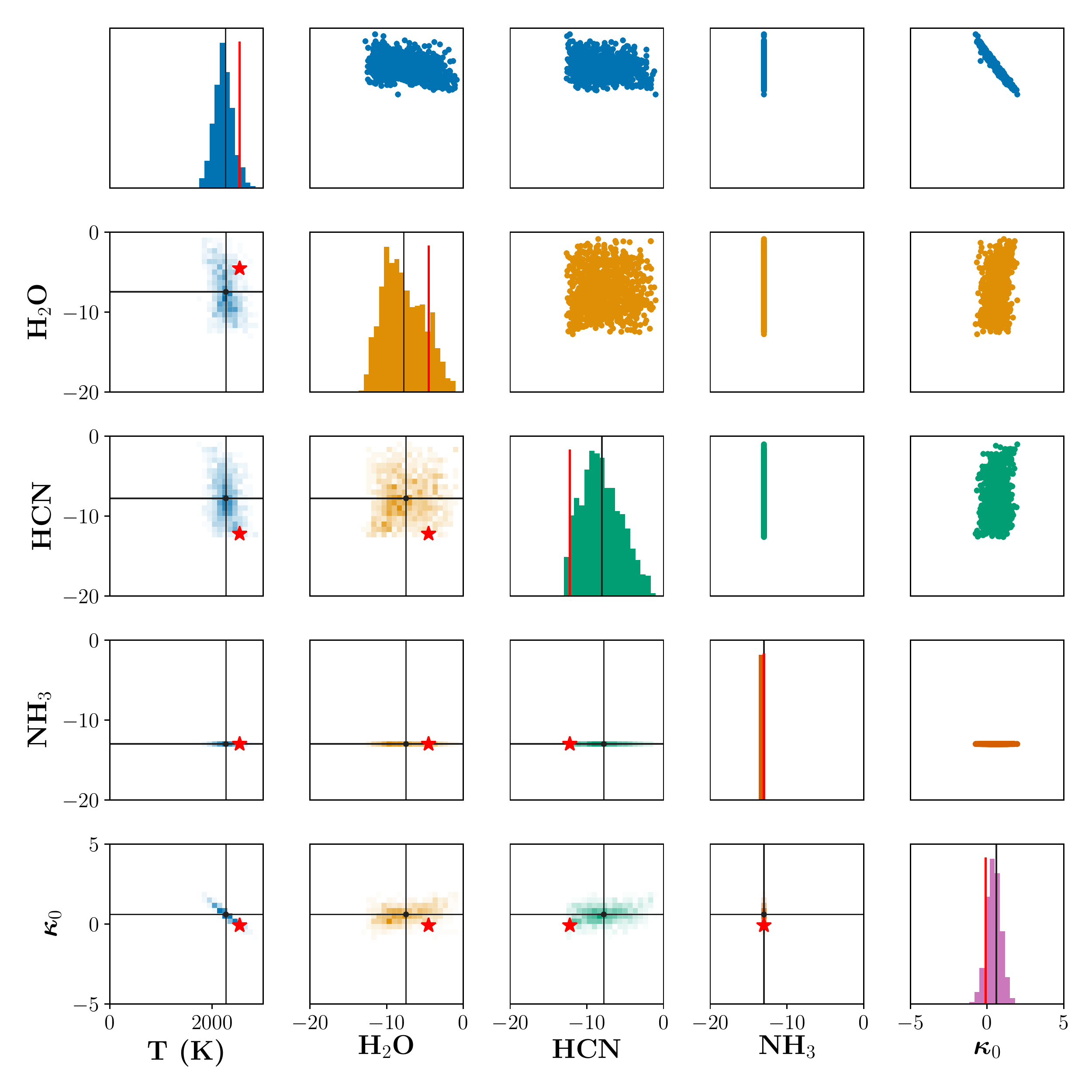}%
  \label{fig:rf_205}%
  }\qquad
  \subfloat[\texttt{plan-net} Ensemble]{%
  \includegraphics[width=.35\textwidth]{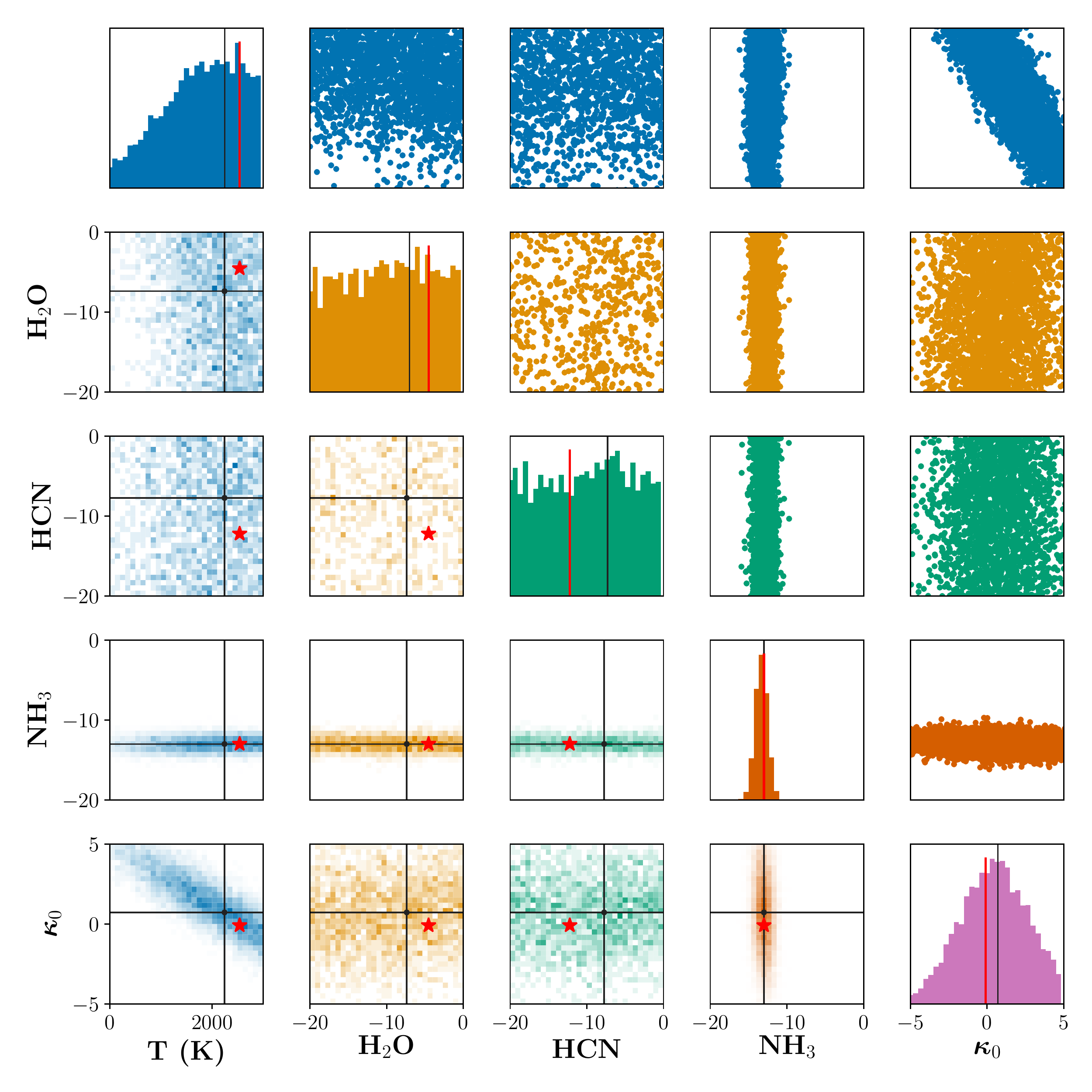}%
  \label{fig:bnn_ens205}
  }\qquad
  \subfloat[Test Spectrum]{%
  \includegraphics[width=.35\textwidth]{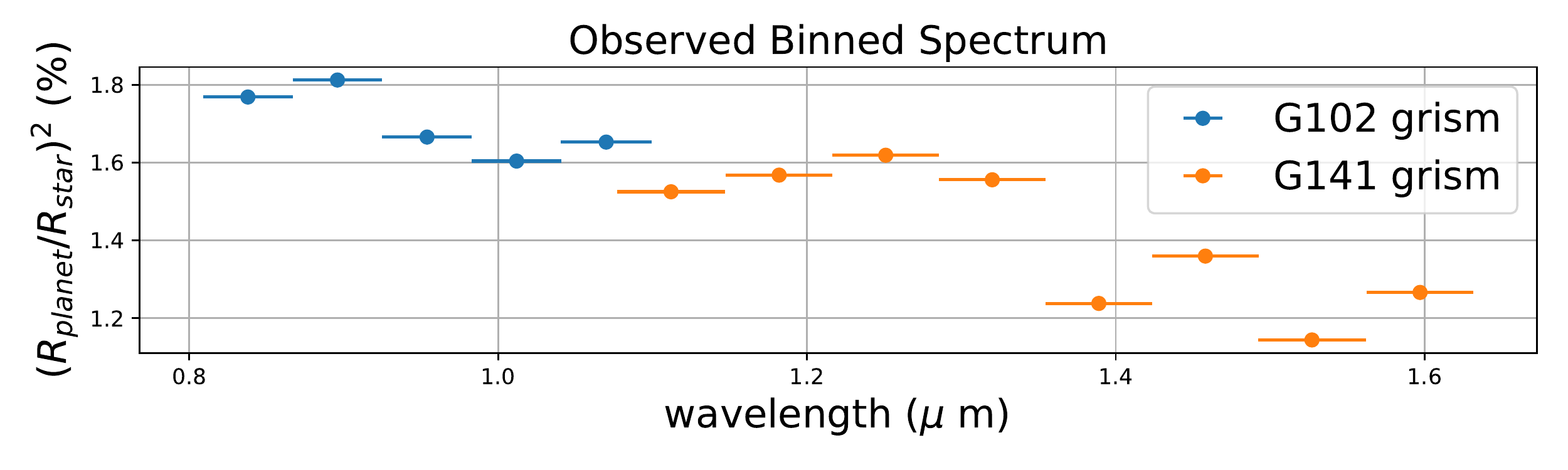}%
  \label{fig:spectra205}
  }\qquad
    \caption{Test planet 2: an example taken from the test set, where both models retrieve parameters close to the true labels, as denoted by the red stars. However, like in Figure \ref{fig:plan_1}, the random forest demonstrates highly confident posteriors, where it may not be appropriate. Figure \ref{fig:spectra205} gives the observed input spectrum, where the binning is given by Table 3 in \cite{KreidbergEtal2015apjWASP12bWater}. Each spectral coverage of the wavelengths is given by two grisms indicated in the legend.}\label{fig:205}
\end{figure}

\begin{figure}[h]
    \centering
    \subfloat[Random Forest]{
\includegraphics[width=.35\textwidth]{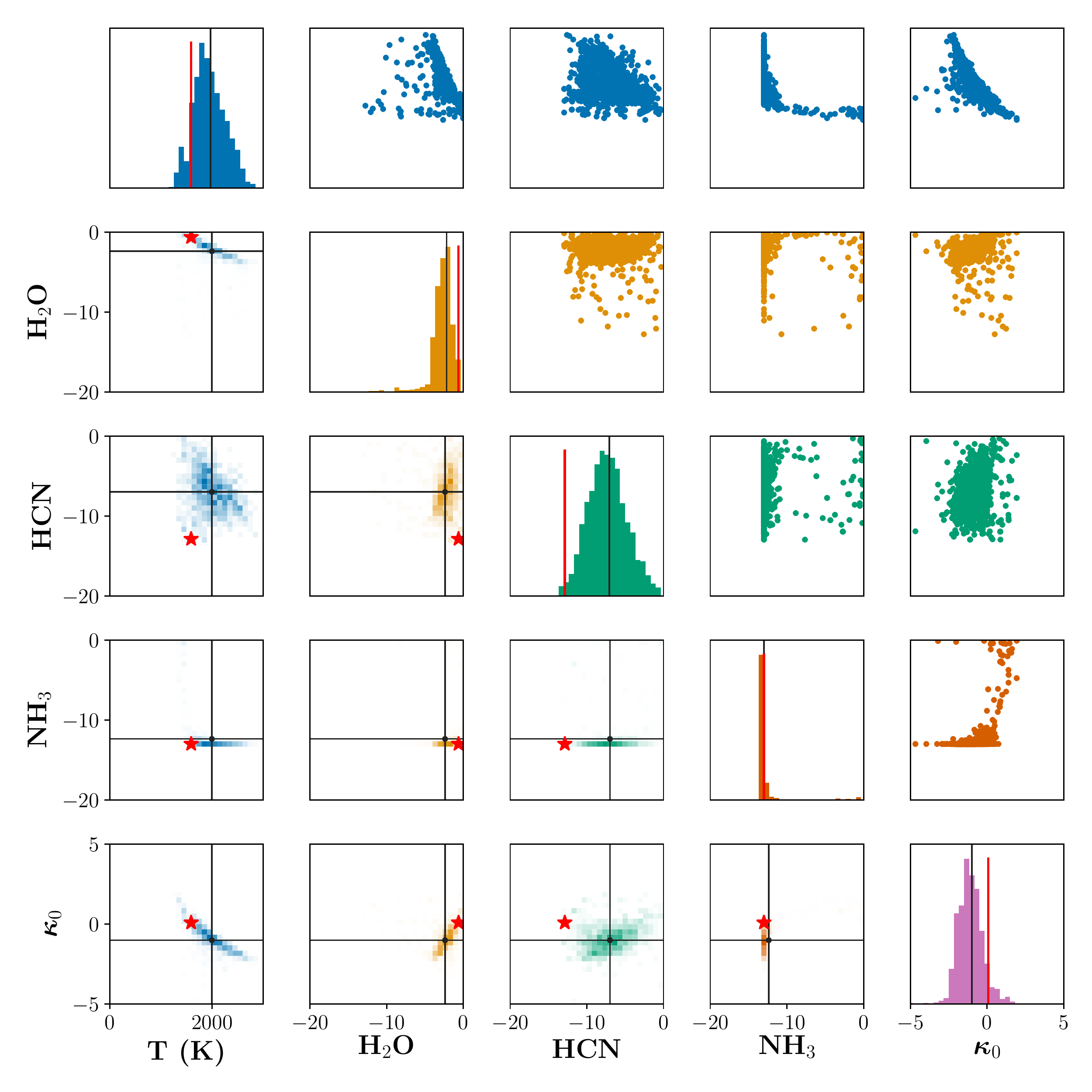}%
  \label{fig:rf_22}%
  }\qquad
  \subfloat[\texttt{plan-net} Ensemble]{%
  \includegraphics[width=.35\textwidth]{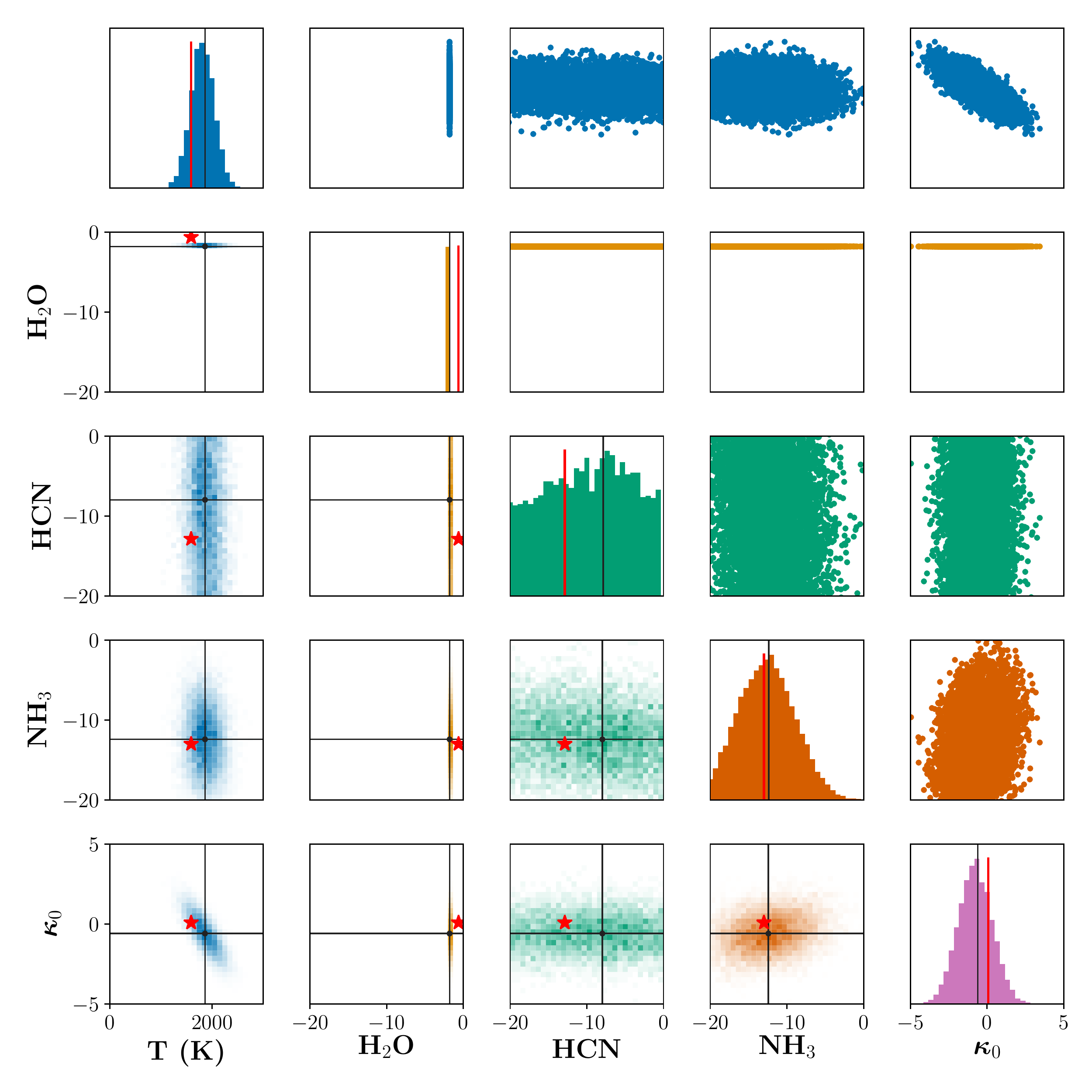}%
  \label{fig:bnn_ens205}
  }\qquad
  \subfloat[Test Spectrum]{%
  \includegraphics[width=.35\textwidth]{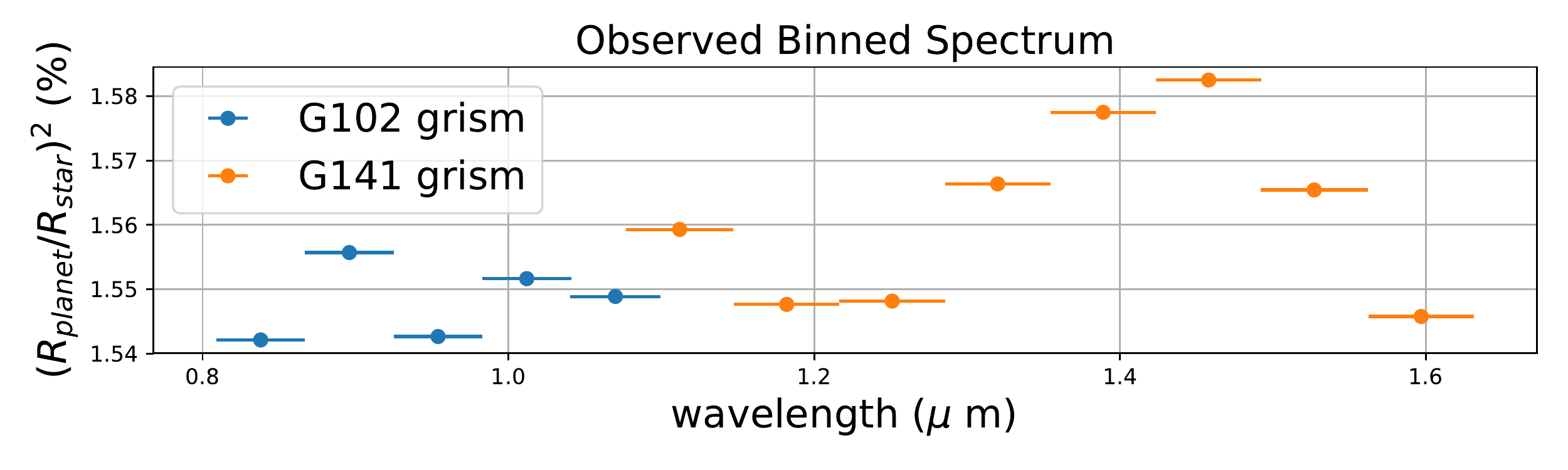}%
  \label{fig:spectra22}
  }\qquad
    \caption{Test planet 3: an example taken from the test set where the H$_2$O abundance is high, allowing it to be tightly constrained. Note that \texttt{plan-net} is unable to constrain the HCN abundance, whereas the RF makes a confident over-prediction; with an abundance of $< 10^{-10}$, HCN would be difficult to constrain for traditional methods \citep{MacDonaldMadhusudhan2017apjlNitrogenChemistryHotJupiters}. Figure \ref{fig:spectra22} gives the observed input spectrum, where the binning is given by Table 3 in \cite{KreidbergEtal2015apjWASP12bWater}. Each spectral coverage of the wavelengths is given by two grisms indicated in the legend.}\label{fig:22}
\end{figure}

\end{appendix}
\end{document}